# Photocurrent and Photoconductance Properties of a GaAs Nanowire

S. Thunich[1], L. Prechtel[1], D. Spirkoska[1], G. Abstreiter[1], A. Fontcuberta i Morral[1,2], and A. W. Holleitner[1,*]

1) Walter Schottky Institut and Physik Department, Technische Universität München, Am Coulombwall 3, D-85748 Garching, Germany.

2) Laboratoire des Matériaux Semiconducteurs, Institut des Matériaux, Ecole Polytechnique Fédérale de Lausanne, CH-1015 Lausanne, Switzerland.

We report on photocurrent and photoconductance processes in a freely suspended p-doped single GaAs nanowire. The nanowires are grown by molecular beam epitaxy (MBE), and they are electrically contacted by a focused ion beam (FIB) deposition technique. The observed photocurrent is generated at the Schottky contacts between the nanowire and metal source-drain electrodes, while the observed photoconductance signal can be explained by a photogating effect induced by optically generated charge carriers located at the surface of the nanowire. Both optoelectronic effects are sensitive to the polarization of the exciting laser field, enabling polarization dependent photodetectors.

*Corresponding author: holleitner@wsi.tum.de

PACS 78.67.-n, 73.21.Hb, 85.60.-q

Semiconductor nanowires have attracted considerable attention for the past few years because of their compelling electronic, mechanical and optical properties.[1]-[10] A very suitable and versatile technique for nanowire growth is the direct synthesis on a substrate.[11]-[14] The fabrication of III-V semiconductor nanowire based devices by such a bottom-up approach ensures the rational use of materials as the nanowires can be obtained in principle on any substrate. Here, we report on the optoelectronic properties of photodetectors based on single p-doped GaAs nanowires grown by molecular beam epitaxy with the so-called vapor-liquid-solid method using Ga droplets as self-catalysts.[15]-[17] The nanowires are electrically contacted by metal electrodes using a focused ion beam (FIB) deposition technique.[18],[19] We experimentally identify two dominating optoelectronic processes in the metal-GaAs nanowire-metal photodetectors by measuring the optoelectronic response of the devices at vacuum as a function of source-drain voltage, modulation frequency, photon energy, spatial coordinate of the optical excitation, and the polarization of the laser field. On the one hand, there is a photocurrent generated at the Schottky contacts between the GaAs nanowires and the metal source-drain electrodes, as recently shown for CdS nanowires.[8] On the other hand, we observe a photoconductance effect, when illuminating the GaAs nanowire far away from the contacts. We interpret the photoconductance effect to arise from band bending effects caused by surface states on the nanowire surface. There, internal electric fields separate the optically excited electron-hole pairs, and in turn, trapped excess electrons act as a negative gating voltage on the p-doped nanowires' surface (photogating effect). At the same time, the optically excited free excess holes raise the Fermi-energy of the hole gas within the nanowires (photodoping effect). Both effects can raise the conductance of the semiconductor circuits.[20],[21] We demonstrate that both photocurrent and photoconductance effects are sensitive to the orientation of linear polarized light. The photoconductance (-current) varies by ~35 % (~15 %) for the photon polarization being parallel or perpendicular to the direction of the GaAs nanowires. Hereby, the metal-GaAs nanowire-metal circuits act as polarization-sensitive photodetectors,[2] which can be integrated into electronic circuits by the FIB-deposition technique in a very versatile way.

Starting point are GaAs nanowires which are grown by molecular beam epitaxy on a $SiO_2$ covered (111)-oriented GaAs substrate as presented previously (growth rate ~0.25 Å/s, $As_4$ partial pressure ~2 · $10^{-6}$ mbar, growth temperature of ~630 °C, and 7 rpm rotation).[17] The nanowires with {110}-facets have a length (diameter) of ~10 μm (~140 nm) and they are doped type p with a carrier concentration of ~$10^{18}$ $cm^{-3}$ by adding a silicon flux during the growth.[17] We present results of two independent samples with optoelectronic characteristics which are typical for the measured p-doped nanowires. The scanning electron micrograph (SEM) image in Fig. 1(a) shows the photodetector circuit with such a p-type GaAs nanowire bridging two gold pads. The pads are first evaporated on top of a $SiO_2$ chip utilizing a shadow mask, and they act as source-drain electrodes for the nanowire. The $SiO_2$ substrate in between the pads is milled away to a depth of a few hundred nanometers by the help of a FIB microscope.[18] Following the technique described in Ref.[19], the particular nanowire is positioned at the shown location by using mechanical manipulators within the FIB recipient. The nanowire is attached to the source-drain electrodes by a FIB induced deposition of carbon utilizing phenanthren [$C_{14}H_{10}$] as a precursor gas [triangles in Fig. 1(a)]. The circuit in Fig. 1(a) features three positions I, II, and III. At position I and III, Schottky contacts between the nanowire and the gold electrodes can be assumed. At position III, however, the nanowire is covered by an opaque carbon layer with a thickness of ~350 nm to cover the underlying Schottky contact. At position II, the nanowire is freely suspended. Hereby, the circuit allows locally addressing the optoelectronic processes in the middle of the nanowires (position II) independently from the ones occurring at the Schottky contact (positions I).

The nanowire devices are mounted in a vacuum chamber at a pressure of about $10^{-5}$ mbar. Optical excitation occurs by focusing the light of a mode-locked titanium:sapphire laser with a repetition rate of 76 MHz through the objective of a microscope onto the nanowire circuits. The laser can be tuned in the energy range 1.24 eV < $E_{PHOTON}$ < 1.77 eV. With a spot diameter of ~2 μm the light intensity $I_{OPT}$ is of the order of 100 W/cm$^2$ for all $E_{PHOTON}$. Generally, we chop the laser at a frequency $f_{CHOP}$. For the optoelectronic measurements discussed below, the resulting current $I_{PHOTO}$ = $I_{ON}(\lambda_{PHOTON}, f_{CHOP})$- $I_{OFF}(\lambda_{PHOTON}, f_{CHOP})$ across the sample with the laser being "ON" or "OFF", respectively, is amplified by a current-voltage converter and detected with a lock-in amplifier utilizing the reference signal provided by the chopper.[21]

We acquire photocurrent images of the nanowire circuits by recording the change of $I_{PHOTO}$ at a finite source-drain bias $V_{SD}$, when the laser spot is laterally scanned across the samples.[20] Then, $I_{PHOTO}$ = $I_{PHOTO}(x, y)$ is plotted using a linear false color scale as a function of the coordinates $x$ and $y$. For all $V_{SD}$, we observe a dominating contribution of |$I_{PHOTO}$| at position I compared to |$I_{PHOTO}$| at position II [Fig. 1(b)]. We also detect $I_{PHOTO}$ as a function of $V_{SD}$ at positions I and II [Fig. 1(c)]. At position I, we observe a finite photocurrent $I_0$ at $V_{SD}$ = 0 V and a finite photovoltage $V_0$ at $I_{PHOTO}$ = 0 A. Hereby, we confirm recent reports that a Schottky contact between a semiconductor nanowire and a metal pad can give rise to both a photocurrent and a photovoltage.[8] In this process, the electron-hole pairs, which are locally created by photoexcitation in the p-doped GaAs nanowires, are separated due to the local built-in electric field at the Schottky contacts, and a maximum (minimum) photocurrent signal can be detected when the illuminated Schottky contact is reverse-biased (forward-biased). We note that at position I, $I_{PHOTO}$ saturates at a source-drain voltage $|V_{SAT}| \sim 0.5$ V, and that both $V_{SAT}$ and $I_{PHOTO}(V_{SAT})$ depend on the laser intensity, as expected for a photocurrent process [data not shown].[22] At position II, however, we observe only a negligible photocurrent and photovoltage signal at zero $V_{SD}$ [Fig. 1(c)]. Here, we find that $I_{PHOTO}$ linearly depends on $V_{SD}$ up to ± 1.5 V [data not shown]. We interpret the observations at position II such that the optoelectronic response of the nanowire far away from the Schottky contacts is dominated by a photoconductance process, which will be discussed in the following.

Fig. 2(a) depicts $|I_{PHOTO}|$ of a nanowire device as a function of $E_{PHOTON}$ for a laser excitation at positions I and II. The dashed line in Fig. 2(a) indicates the band gap energy $E_{GaAs} \approx 1.42$ eV of GaAs at room temperature,[23] while the noise level in the set-up without laser excitation is below ~2 pA. Since we detect a finite $|I_{PHOTO}|$ for $E_{PHOTON} < E_{GaAs}$ for all positions on the nanowires, we conclude that there are parts of the nanowires, where an electric field gives rise to a Franz-Keldysh effect of the absorption energy,[24]-[27] and in addition, where surface states are optically excited. We infer that the electric fields arise independently of the Schottky contacts within the nanowire circuits. Instead, we assume that the electric field can be attributed to the effective mid-gap Fermi-level pinning of the oxidized {110}-facets of the p-doped GaAs with respect to the electrostatic potential of the holes in the center of the nanowires.[28],[29] The expected field strength can be estimated by the band gap energy of GaAs and the lateral depletion length of typically $l_{DEP}$ ~ 20 nm.[30] The resulting field strength $F \sim 0.5 \cdot E_{GaAs} / (e \cdot l_{DEP}) \sim 10^7$ V/m is sufficient to ionize optically created excitons at the surface region of the nanowires.[24],[25] Separated charge carriers can contribute twofold to the photoconductance of the nanowires. On the one hand, the electrons can drift to the surface of the p-doped nanowires, acting as a local negative gating voltage (photogating effect or normal surface photovoltage). Hereby, the conductive part of the nanowire is widened. On the other hand, the holes increase the hole density within the nanowires (photodoping effect).[31] Both effects can result in an optically increased conductance of the nanowire circuits.[20],[21]

The observed photoconductance and -current signals depend on the polarization of the exciting laser field for both positions I and II. The corresponding signals follow a $\cos\phi$-like dependence, where $\phi$ is the angle between the wire and the light polarization. This observation is in agreement with recent reports that the nanowire geometry strongly affects the polarization of emitted or absorbed photons.[2] We observe that the polarization ratio $\rho = (I_{PHOTO}^{\parallel} - I_{PHOTO}^{\perp}) / (I_{PHOTO}^{\parallel} + I_{PHOTO}^{\perp})$ of $I_{PHOTO}$ with parallel ($I_{PHOTO}^{\parallel}$) and perpendicular ($I_{PHOTO}^{\perp}$) polarization orientation is larger at position II (~ 35 %) than close to the Schottky contact of position I (~ 15 %). Hereby, the data in Fig. 2(b) demonstrate that one can simply enhance $\rho$ by the described FIB-deposition technique, since the technique allows suppressing the optoelectronic response of the nanowires due to the Schottky contacts by covering the contacts by an opaque carbon layer [position III in Fig. 1(a) and (b)].

In order to test the response time of the metal-nanowire-metal photodetector, we measure $I_{PHOTO}$ of the devices as a function of $f_{CHOP}$ [Fig. 2(c)]. For both positions I and II, we do not detect any dependence of $I_{PHOTO}$ on $f_{CHOP}$ for frequencies between 0.1 and 6 kHz. This reveals that the effects causing the photocurrent (position I) and the photoconductance (position II) occur on a time-scale shorter than ~ (6 kHz)$^{-1}$ = 167 µs. Recently, we reported on photoconductance processes in both submicron channels and quantum wires, which are fabricated in AlGaAs/GaAs heterostructures by etching techniques.[20],[21] There, we found response times of the photoconductance in the range of several milliseconds. The occurrence of longer response times in such AlGaAs/GaAs heterostructure based devices is consistent with the fact that there, spatially separated charge carriers can be located in separated layers of the heterostructure with energy barriers in between. In the case of nanowires, the electrons and the holes are separated by a maximum distance of the depletion length. Hereby, we explain the fact, that we measure relatively faster response times in the metal-GaAs nanowires-metal detectors.

In summary, we present spatially and spectrally resolved optoelectronic measurements on freely suspended p-doped GaAs nanowires. We interpret the results in terms of the combination of a photogating and a photodoping effect, i. e. the consequence of the spatial separation of holes and electrons, in combination with a photocurrent effect which is generated at the Schottky barriers between the GaAs nanowires and the metal electrodes. We demonstrate that both effects allow building a polarization dependent photodetector with a response time faster than ~200 µs.

We gratefully acknowledge financial support by the DFG (Ho 3324/4), the German excellence initiative via the "Nanosystems Initiative Munich (NIM)", Marie Curie Excellence Grant SENFED, the Swiss National Science Foundation for the project "Catalyst-free direct doping of MBE grown III-V nanowires", as well as the "Center for NanoScience" (CeNS) in Munich.

**Fig. 1.** (Color online) (a) Scanning electron microscope (SEM) image of a freely suspended p-doped GaAs nanowire bridging two gold electrodes. Positions I and III denote the Schottky contacts, while position II highlights the center part of the nanowire. The Schottky contact at position III is covered by an opaque carbon deposit. (b) Photocurrent contour plots of $|I_{PHOTO}|$ as a function of the spatial coordinates ($T_{BATH}$ ~ 296 K, $P_{LASER}$ ~ 60 W/cm$^2$, $f_{CHOP}$ = 1028 Hz, $\lambda$ = 780 nm, $V_{SD}$ = +1V). (c) $I_{PHOTON}$ as function of $V_{SD}$ for positions I and II ($T_{BATH}$ ~ 296 K, $P_{LASER}$ ~ 30 W/cm$^2$, $f_{CHOP}$ = 930 Hz, $\lambda$ = 730 nm).

**Fig. 2.** (Color online) (a) $|I_{PHOTO}|$ as a function of $E_{PHOTON}$ for positions I (circles) and II (squares) as defined in Fig. 1(a) ($V_{SD}$ = -1 V, $T_{BATH}$ ~ 296 K, $P_{LASER}$ ~ 100 W/cm$^2$, $f_{CHOP}$ = 930 Hz). (b) $|I_{PHOTO}|$ as a function of the angle $\phi$ between the linearly polarized laser field and the orientation of the nanowire for positions I (circles) and II (squares) ($V_{SD}$ = +0.5 V, $T_{BATH}$ ~ 296 K, $P_{LASER}$ ~ 60 W/cm$^2$, $f_{CHOP}$ = 930 Hz, $\lambda$ = 800 nm). (c) $|I_{PHOTO}|$ as a function of $f_{CHOP}$ for positions I (circles) and II (squares) ($V_{SD}$ = -1 V, $T_{BATH}$ ~ 296 K, $P_{LASER}$ ~ 60 W/cm$^2$, $\lambda$ = 780 nm).

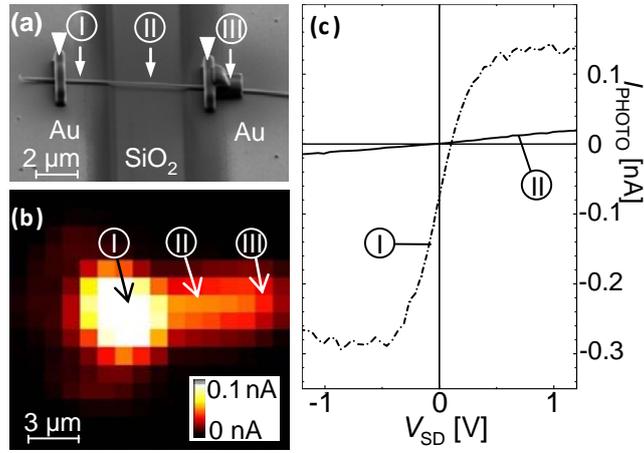

Figure 1

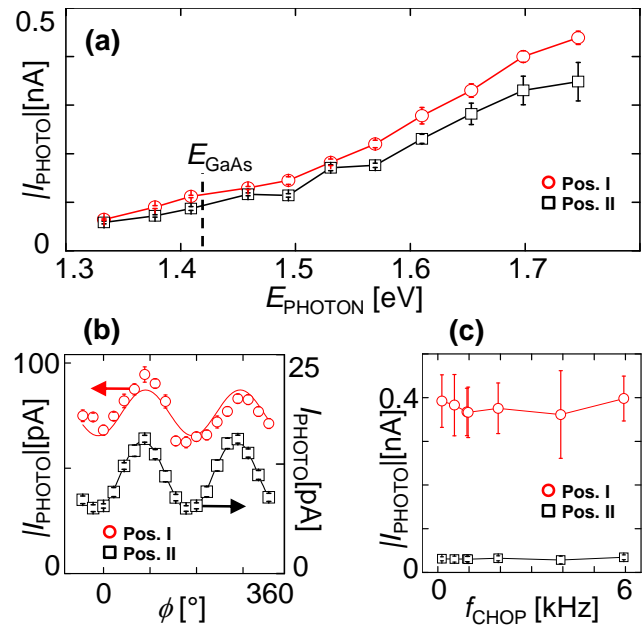

Figure 2